# A Teflon-based system for applying multidirectional voltages to lipid bilayers as a novel platform for membrane proteins

Maki Komiya[1, †], Kensaku Kanomata[2, †], Ryo Yokota[1], Yusuke Tsuneta[1], Madoka Sato[1], Daichi Yamaura[1], Daisuke Tadaki[1], Teng Ma[3], Hideaki Yamamoto[3], Yuzuru Tozawa[4], Albert Martí[5], Jordi Madrenas[5], Shigeru Kubota[2], Fumihiko Hirose[2], Michio Niwano[6], and Ayumi Hirano-Iwata[1, 3, *]

[1] Laboratory for Nanoelectronics and Spintronics, Research Institute of Electrical Communication, Tohoku University, 2-1-1 Katahira, Aoba-ku, Sendai-shi, Miyagi 980-8577, Japan

[2] Graduate School of Science and Engineering, Yamagata University, 4-3-16 Jonan, Yonezawa-shi, Yamagata 992-8510, Japan

[3] Advanced Institute for Materials Research, Tohoku University, 2-1-1 Katahira, Aoba-ku, Sendai-shi, Miyagi 980-8577, Japan

[4] Graduate School of Science and Engineering, Saitama University, 255 Shimo-Okubo, Sakura-ku, Saitama-shi, Saitama 338-8570, Japan

[5] Department of Electronics Engineering, Universitat Politècnica de Catalunya, Campus Nord, Building C4. C. Jordi Girona, 1-3, 08034 Barcelona, Catalunya, Spain

[6] Kansei Fukushi Research Institute, Tohoku Fukushi University, 6-149-1 Kunimi-ga-oka, Aoba-ku, Sendai-shi, Miyagi 989-3201, Japan

†These authors contributed equally to this work.
*Authors to whom correspondence should be addressed.
Email: ayumi.hirano.a5@tohoku.ac.jp

**Abstract**

Artificial bilayer lipid membranes (BLMs), along with patch-clamped membranes, are frequently used for functional analyses of membrane proteins. In both methods, the electric properties of membranes are characterized by only one parameter, namely, transmembrane potential. Here the construction of a novel BLM system was reported, in which membrane voltages can be controlled in a lateral direction in addition to conventional transmembrane direction. A microaperture was fabricated in a Teflon film and Ti electrodes were evaporated around the aperture. BLMs were reproducibly formed in the aperture without being affected by the presence of the electrodes. The application of a lateral voltage induced no significant changes in the electric properties of the BLMs, such as baseline current, transmembrane resistance, and transmembrane capacitance. In contrast, lateral voltages clearly affected the activities of biological ion channels, suggesting that the lateral voltage might be a useful parameter for analyzing channel activities. The present Teflon-based system in which multidirectional voltages can be applied to BLMs represent a promising platform for the analysis of underlying functional properties of membrane proteins.

**Introduction**

The biological cell membrane that separates the cell from its external environment is composed of a bilayer lipid membrane (BLM) and membrane proteins. The BLM is a self-assembled structure comprised of phospholipid molecules that stabilize the membrane through hydrophilic and hydrophobic interactions. It provides an environment that permits membrane proteins to express their functions. There are two common methods for the electric functional analysis of membrane proteins, especially ion channels; the patch clamp technique for proteins that are in cell membranes[1–3] and the reconstitution of ion channels in artificially formed BLMs[4–10]. Although recordings from patch-clamp preparations and channel-incorporated BLMs provide basic information concerning ion channel biophysics, only one electric parameter can be measured in either type of membrane, i.e., transmembrane potential. However, BLM is a three-dimensional material with nanometer-thickness and application of voltages in other directions might be another parameter for the functional analysis of ion channels. We report herein on the construction of a BLM system in which we are able to regulate membrane voltages in a lateral direction in addition to conventional transmembrane direction (Figure 1). Teflon films, which are traditionally used for suspending BLMs[4,11–14], were used as a base material, on which Ti electrodes were deposited. The resulting structure, referred to as a Teflon-based chip, was sufficiently stable to permit the formation of BLMs. Using BLMs in the Teflon-based chip, we investigated the effects of lateral voltages on background membrane properties, such as transmembrane currents, resistance, and capacitance, together with the effects of lateral voltages on the activities of biological ion channels. We show that ion channel activities can be regulated by the lateral voltages while the application of the lateral voltages induces no significant changes in electric properties of BLMs. Therefore, the present system has the potential to serve as a novel platform for revealing the fundamental and functional properties of membrane proteins.

**Results and Discussion**

Since early BLM studies, Teflon films have been used as a septum for suspending BLMs[4,11–14], owing to their chemical stability and good electric properties, such as high resistivity, low dielectric constant, and low dielectric loss.[15] We fabricated novel Teflon-based chips for applying multidirectional voltages in BLMs by utilizing Teflon films as a base material. Microapertures were first produced in Teflon films by an electrical spark, and then Ti electrodes and $SiO_2$ layers were deposited around the apertures (Figure 2). Although numerous crack-like lines were formed on the Teflon-based chips after the deposition of the $SiO_2$ layers, the Teflon-based chip showed a resistance of ~2 TΩ when immersed in a buffer solution, confirming that the Teflon-based chip had a high insulating property (Figure S1). After silanizing the surface of the $SiO_2$ layer with PFDS, BLM formation in the Teflon-based chips was investigated. BLMs with resistances of >10 GΩ and >100 GΩ were formed in the Teflon-based chips with success probabilities of 83% and 70%, respectively, based on 81 trials. Thus, BLMs were reproducibly formed in the Teflon-based chips without being affected by the presence of the Ti and $SiO_2$ layers around the apertures.

We next applied lateral voltages to BLMs through the Ti electrodes and investigated the effects of the lateral voltages on transmembrane currents. Figure 3(a) shows current traces recorded at a transmembrane voltage of +100 mV in the presence or absence of a lateral voltage of 3 V. No significant change in either baseline current nor the noise level were induced by the application of lateral voltages (P>0.1, n=33, Student's two-tailed t-test for paired samples). We then investigated the effects of the lateral voltage on transmembrane resistance and capacitance. Without a lateral voltage, the transmembrane resistance was close to 200 GΩ (Figure 3(b)), which was approaching the upper limit of the patch-clamp amplifier in our recording system. When the Ti electrodes on the Teflon-based chip were connected to a DC power source and the lateral voltage was increased from 0 to 4 V, the transmembrane resistance remained constant. Similarly, the application of the lateral voltage induced no significant change in the total capacitance (the sum of

BLM and device capacitance) in the transmembrane direction. Thus, no significant change was induced in the electric properties of the BLMs in terms of baseline current level, transmembrane resistance and capacitance.

To demonstrate the utility of the present BLM system as a novel platform for the functional analysis of membrane proteins, we incorporated biological ion channel proteins into the BLMs, and the effect of lateral voltages on channel activities were then investigated. For this purpose, proteoliposomes were prepared from HEK293T cells that expressed a cardiac voltage-dependent sodium channel (Nav1.5) and added to the *cis* side compartment in order to incorporate the channels into the BLMs that were formed in the Teflon-based chips. Figure 4 shows the channel currents recorded at a transmembrane voltage of +100 mV after a 10 ms prepulse of -200 mV. Such a strong hyperpolarizing prepulse is commonly used as a voltage protocol for the Nav1.5 channel in patch-clamp studies[16,17]. Without lateral voltages, channel activities were infrequently observed. Only three opening events were observed in 23 sweeps of the voltage protocol (Figure 4(a)) and no channel activities were subsequently detected for the following 60 sweeps (Figure 4(b)), indicating that the channel activities had disappeared. We then connected the DC power source to the Ti electrodes on the Teflon-based chip and applied a lateral voltage between the electrodes. When we applied a lateral voltage of 2 V, channel activities were dramatically enhanced, with frequent opening and closing events being detected (Figure 4(c)). Channel activities were elicited by every sweep of the voltage protocol which was repeated 17 times. In contrast, when the lateral voltage was switched off, channel activities disappeared completely (Figure 4(d)). We next omitted the hyperpolarizing prepulse from the voltage protocol and investigated the channel activities at a constant transmembrane voltage of +100 mV. Figure 5(a) shows example traces of the channel currents, which were recorded under the application of the lateral voltage. The observed single-channel conductance (31 pS) was consistent with an estimated value (32 pS) that was calculated for the present $Na^+$ concentration gradient (149.2 mM/5 mM) based on a reported conductance (17 pS)[17] of the Nav1.5 channel according to the Nernst equation. Switching off the lateral voltage resulted in the disappearance of the channel activities (Figure 5(b)). When we applied the lateral voltage again, the channel activities revived (Figure 5(c)), suggesting that the effect of the lateral voltage was reversible. Considering the fact that BLMs without the channels exhibited no perturbation currents, even when a much higher lateral voltage (3 V) was applied (Figure 3(a)), these results indicate that ion channel activities were affected by the application of a lateral voltage. Similar activation effects of lateral voltages were observed for 83% of the BLMs that exhibited channel activities (n = 6). The possibility of current leakage through the aqueous solution was rejected, since observed lateral currents were less than 0.2 pA at a lateral applied voltage of 4 V due to the very high resistance of the BLM and $SiO_2$ layer (Figure S1).

**Conclusion**

In conclusion, we report on the development of a novel BLM system in which the membrane potential can be controlled in the lateral direction to BLMs in addition to the conventional transmembrane direction. By utilizing artificial BLM systems, it was possible to position electrodes inside lipid bilayer structures through the deposition of metal electrodes on a Teflon film to which the BLMs were suspended. The above findings suggest that ion channel activities are reversibly regulated by a lateral voltage. Although it remains to be elucidated how the lateral voltage is related to channel openings, the application of the lateral voltage has the potential to open the inactivated ion channels. This would be useful for the analysis of fast-inactivating ion channels that were difficult to be evaluated based on the application of conventional transmembrane voltages. Considering that inserting electrodes between very thin lipid monolayers of living cell membranes is not easy in the patch-clamp method, the present BLM system represents a promising tool for use in functional analyses of membrane proteins, including ion channels.

**Experimental Section**

**Fabrication of Teflon-based chips:** Teflon films with thicknesses in the range of 12 to 15 μm (YSI Inc. High Sensitivity Membrane Kit) were cut into rectangular sheets with a size of 32 × 40 mm$^2$. Small, circular apertures (70~180 μm) across which BLMs were formed were produced in the Teflon sheet by passing an electrical spark generated by an automobile ignition coil.[14] Ti films with a thickness of 200 nm were then deposited on the sheet through a metal mask by using an electron beam evaporator (ANELVA VT-43N) with a Ti deposition rate of 0.15 nm/s. The temperature of the Teflon sheet was maintained at room temperature (21 ºC) during the deposition. The Ti films were used as metal electrodes to apply a lateral voltage inside a BLM. To electrically isolate the Ti electrodes from buffer solutions, $SiO_2$ films with a thickness of 100 nm were deposited on the Ti films with the electron beam evaporator at room temperature at a $SiO_2$ deposition rate of 0.2 nm/s. The microstructure around the aperture was observed using a field emission scanning electron microscope (FE-SEM) (Hitachi High-Technologies SU-8000).

**Preparation of Proteoliposomes:** HEK293T cells were maintained in DMEM (1×) + GlutaMAX™-I (gibco) supplemented with 10 % FBS (gibco) under 5% $CO_2$ at 37°C. HEK293T cells at a 70-80% confluency were transfected with Nav1.5 DNA subcloned in a pCMV6-XL4 vector [Nav1.5(SCN5A) (NM_198056) Human Untagged Clone, ORIGENE] by using PEI MAX (Polysciences, Inc.). More than 24 hours after the transfection, the Nav1.5 channels were extracted from the transfected HEK293T cells as membrane fractions according to procedures described in the previous report[18]. In brief, the cells from 100-mm plates were rinsed with HBSS (gibco) and scraped off into a solution of 200 mM KCl, 33 mM KF, 10 mM EDTA, 50 mM HEPES (pH 7.4 with KOH) plus protease inhibitors (100 μM phenylmethylsulfonyl fluoride, 1 μg/ml pepstatin A, 1 μg/ml leupeptin). The cells were homogenized and spun at 1500 × g for 10 min. The low-speed supernatants were collected and centrifuged at 7734 × g for 30 min. The supernatants were then centrifuged at 157000 × g for 1 h. All steps were performed at 4 °C. The membrane pellets were resuspended in a 4:1 (v/v) mixture of high-$Na^+$ buffer (149.2 mM NaCl, 4.7 mM KCl, 2.5 mM $CaCl_2$, 5 mM HEPES (pH 7.3 wih NaOH)) and glycerol, and stored at 4 °C under nitrogen.

**BLM formation and channel incorporation:** The Teflon-based chips were washed with chloroform, ethanol, and toluene, and then immersed in a 2% (v/v) solution of (tridecafluoro-1,1,2,2-tetrahydrooctyl)-dimethylchlorosilane (PFDS) in super dehydrated toluene at room temperature for 6 hours in a nitrogen-filled glovebox. The BLMs were formed across the aperture in the silanized Teflon-based chip by the folding method, as described in a previous report[12]. In brief, the silanized Teflon-based chips were placed in the middle of a Teflon recording chamber. The Teflon-based chip separated the *cis* and *trans* compartments in the chamber. After washing the chamber with water, ethanol and chloroform, each side of the Teflon-based chip around the aperture was precoated with a thin layer of *n*-hexadecane using a cotton swab. A 1400 μL volume of high-$Na^+$ buffer (149.2 mM NaCl, 4.7 mM KCl, 2.5 mM $CaCl_2$, 5 mM HEPES (pH 7.3 with NaOH)) and low-$Na^+$ buffer (5 mM NaCl, 143.9 mM KCl, 5 mM $MgCl_2$, 5 mM HEPES (pH 7.3 with KOH)), filtered through a cellulose acetate filter (pore size 0.20 μm; Advantec Toyo, Tokyo, Japan), was added to *cis* and *trans* compartment, respectively. This $Na^+$ concentration gradient was applied in order to increase the channel currents. The water level in both compartments was initially set below the aperture. A 30-μL portion of a lipid solution (5 mg/mL, L-α-phosphatidylcholine: L-α-phosphatidyl-ethanolamine: cholesterol = 7: 1: 2 (w/w) in chloroform/*n*-hexane (1:1, v/v)) was then carefully spread on the buffer solution in each compartment. After evaporation of the solvent, a BLM was produced by gradually increasing the water level until it surpassed the aperture. After BLM formation, a 30-60 μL portion of proteoliposomes containing the Nav1.5 channels was added to the solution in the *cis* compartment to incorporate the channels into the BLMs.

**Current recordings:** Recording transmembrane currents were performed with a patch-clamp amplifier (Molecular Devices Axopatch 200B). The current signals were filtered online at 2 kHz with a low-pass Bessel filter, digitized at 25 kHz, and stored online using a data acquisition system (Molecular Devices Digidata 1440A and 1550B using pClamp 10.3 and 10.6, respectively). Transmembrane voltage was defined with respect to the *trans* side, which was held at ground. Transmembrane resistance was calculated based on the transmembrane currents observed at transmembrane voltages of +100 mV and -100 mV. Transmembrane capacitance was measured by applying a ramp transmembrane voltage wave of 1 V/s at a holding potential of 0 mV. A homemade battery-operated DC power source was used to apply a lateral voltage through the connection with the Ti electrodes that were deposited on the Teflon-based chip. The lateral current under the laterally applied voltage was measured by a system sourcemeter[R] (A Tektronix Company KEITHLEY 2636B) which was connected in series with the DC power source and the Teflon-based chip.

**Acknowledgements**

This work was supported by the CREST program of the Japan Science and Technology Agency (JPMJCR14F3), JSPS KAKENHI (19H00846), and the Nation-wide Cooperative Research Project Program of the Research Institute of Electrical Communication at Tohoku University. The authors wish to thank Mr. Shohei Fujii at Kyoto Sangyo University for useful discussion on proteoliposome extraction. We also wish to thank Ms. Yukie Takano at Saitama University for amplifying plasmid DNAs. Some of the equipment used in this research was manufactured by Kento Abe, a technical staff member in the machine shop division of Fundamental Technology Center, Research Institute of Electrical Communication, Tohoku University.

**Conflict of Interest**

The authors declare no conflict of interest.

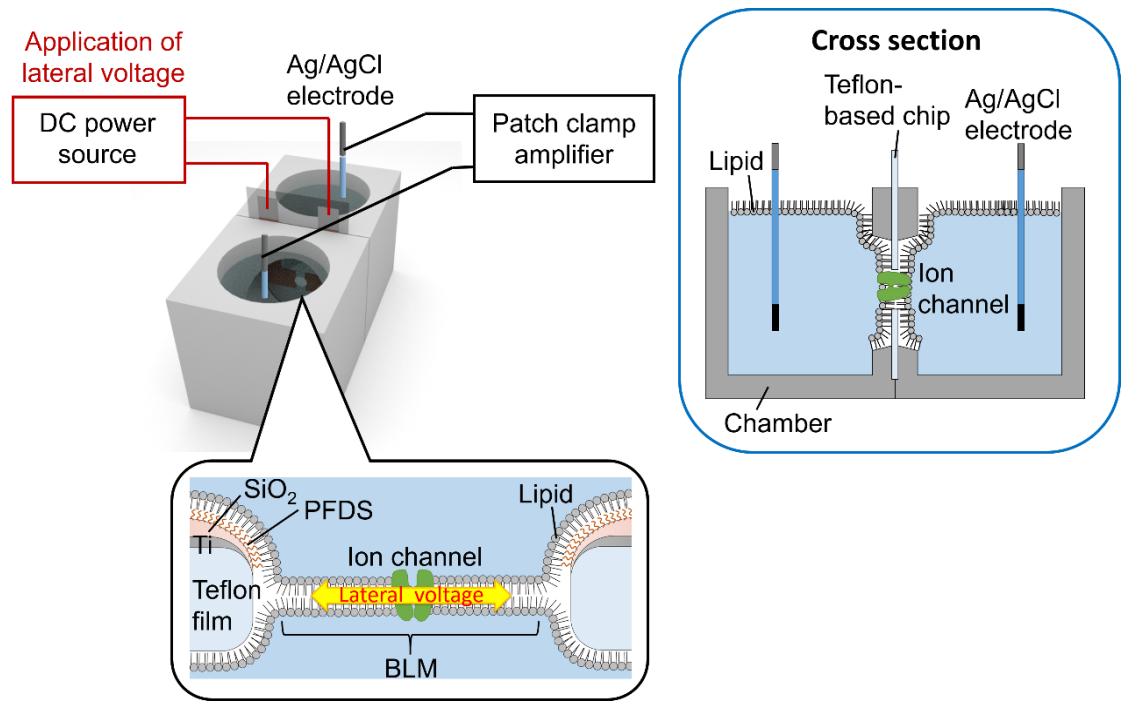

**Figure 1. A schematic illustration of a BLM in a Teflon-based chip with embedded Ti electrodes.**

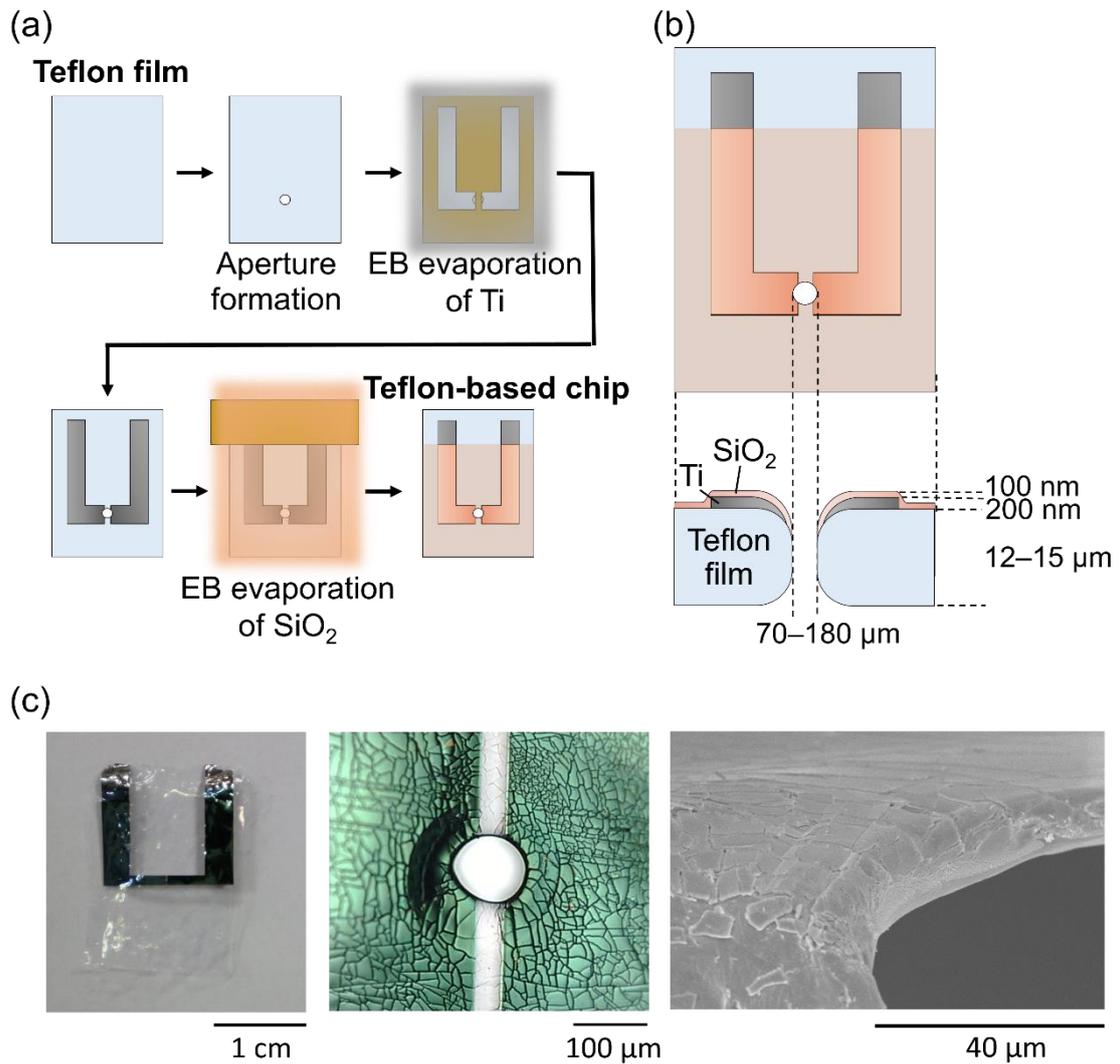

**Figure 2. (a)** Schematics of the procedure for fabricating Teflon-based chips. **(b)** Schematics of a fabricated Teflon-based chip. **(c) (left)** A photograph of a Teflon-based chip having an aperture. **(center)** A photomicrograph of the Teflon-based chip around the aperture. **(right)** An FE-SEM image around the aperture.

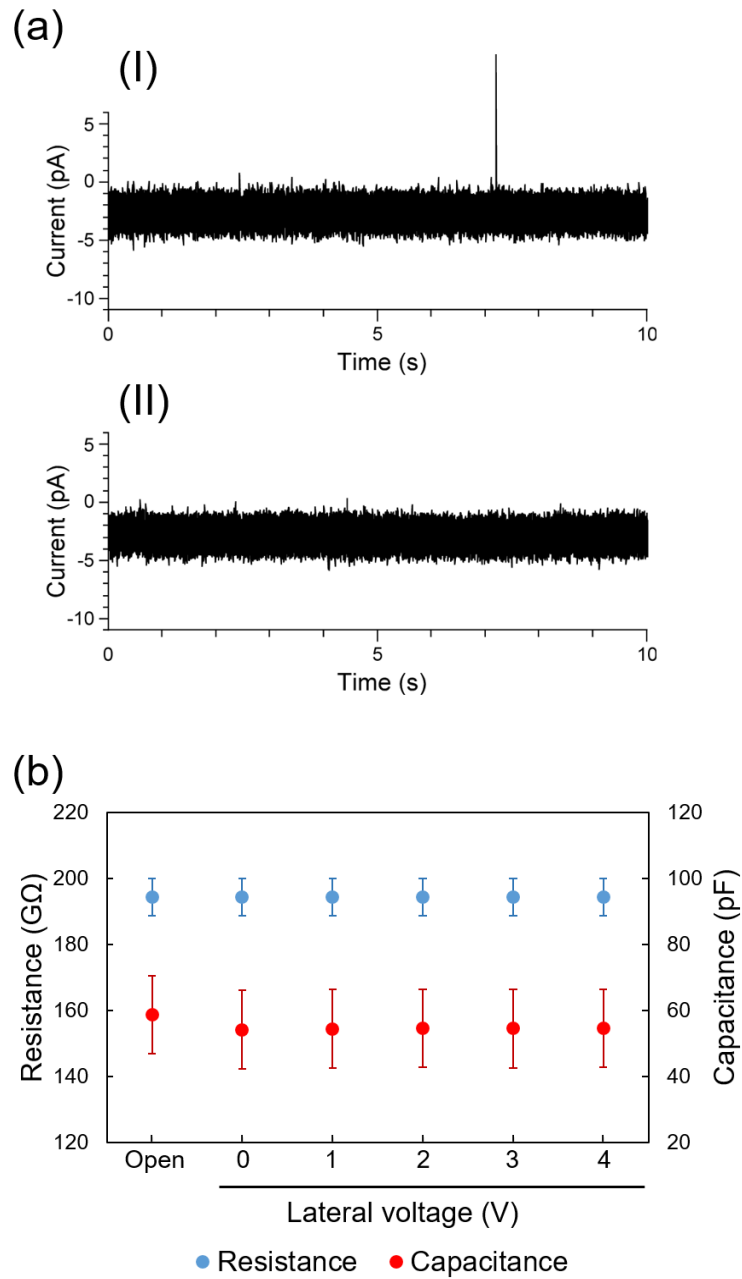

**Figure 3. Effects of lateral voltages on transmembrane electric properties.**
(a) An example of current traces recorded at a transmembrane voltage of +100 mV with a BLM formed in a Teflon-based chip. (I) Without a connection between the Teflon-based chip and a DC power source (open state). (II) With a lateral voltage of 3 V, which was applied by a DC power source. (b) Relationships between lateral voltages and either BLM resistance or total capacitance (sum of BLM and device capacitance) in the transmembrane direction (n = 10). The resistance higher than 200 GΩ was counted as 200 GΩ. The error bar is SEM.

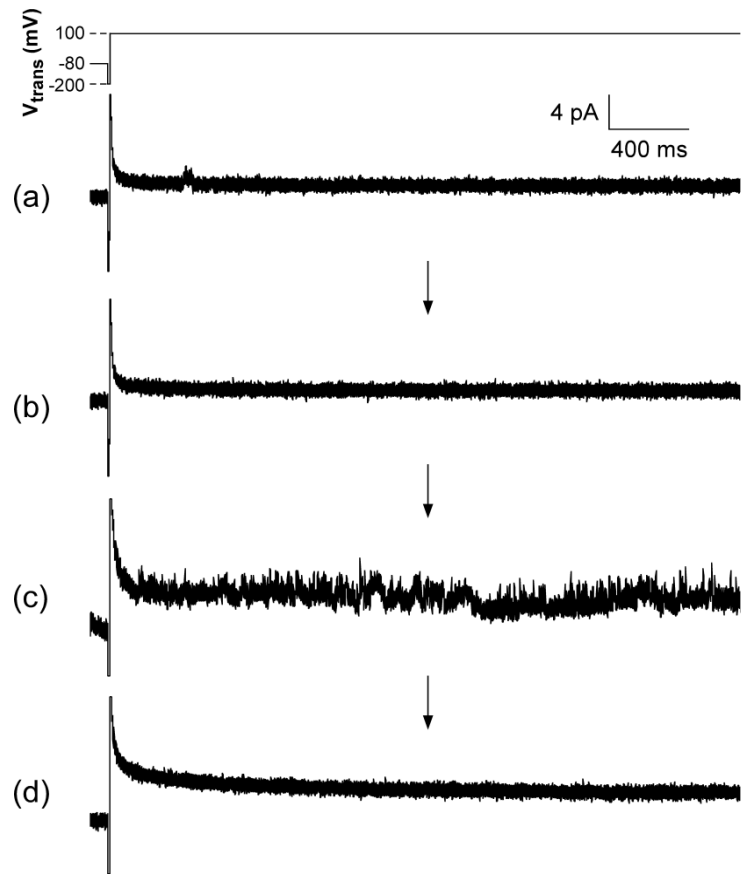

**Figure 4. Effect of lateral voltage on ion channel currents recorded from a BLM formed in a Teflon-based chip.**
Transmembrane currents were recorded after addition of proteoliposomes containing Nav1.5 channels using a transmembrane voltage protocol with a 10 ms step to -200 mV, followed by a step to +100 mV. (a) and (b) Example of current traces that showed (a) a channel opening event and (b) no channel activities in the absence of the lateral voltage. (c) and (d) Example of current traces that were recorded when the lateral voltage of 2 V was (c) switched on and (d) switched off.

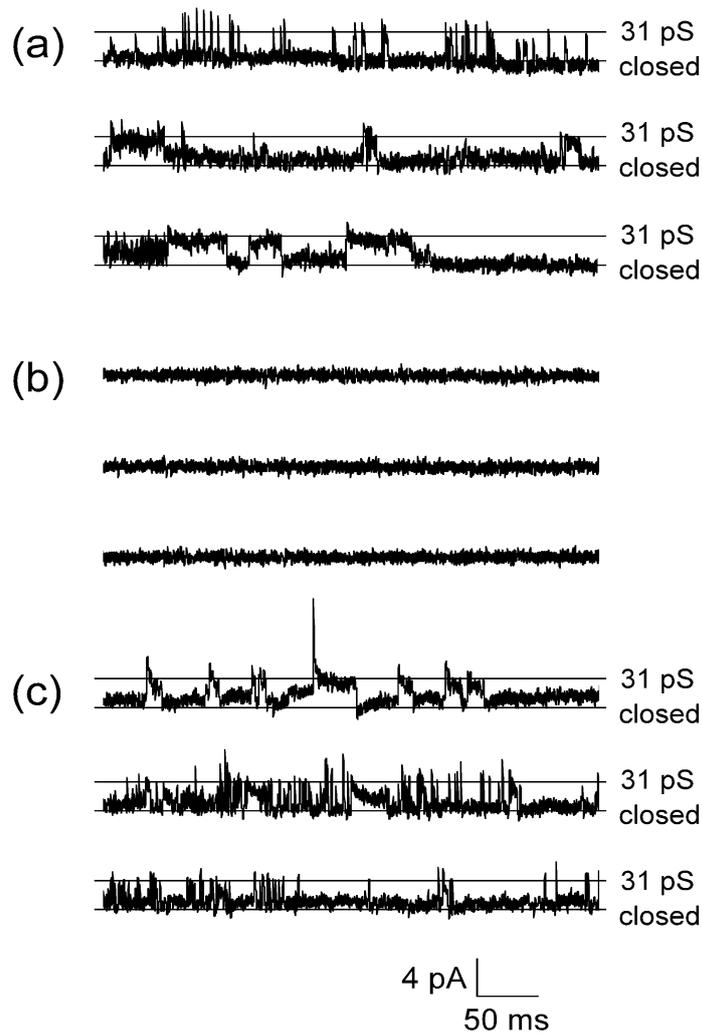

**Figure 5. Effect of a lateral voltage on ion channel currents recorded from a BLM formed in a Teflon-based chip.**
Transmembrane currents were recorded at a transmembrane voltage of +100 mV after the addition of proteoliposomes containing Nav1.5 channels. Example of current traces that were recorded when the lateral voltage of 2 V was (a) switched on, (b) switched off and (c) switched on again.

*Supporting Information*

# A Teflon-based system for applying multidirectional voltages to lipid bilayers as a novel platform for membrane proteins

Maki Komiya[1,†], Kensaku Kanomata[2,†], Ryo Yokota[1], Yusuke Tsuneta[1], Madoka Sato[1], Daichi Yamaura[1], Daisuke Tadaki[1], Teng Ma[3], Hideaki Yamamoto[3], Yuzuru Tozawa[4], Albert Martí[5], Jordi Madrenas[5], Shigeru Kubota[2], Fumihiko Hirose[2], Michio Niwano[6], and Ayumi Hirano-Iwata[1,3,*]

[1] Laboratory for Nanoelectronics and Spintronics, Research Institute of Electrical Communication, Tohoku University, 2-1-1 Katahira, Aoba-ku, Sendai-shi, Miyagi 980-8577, Japan

[2] Graduate School of Science and Engineering, Yamagata University, 4-3-16 Jonan, Yonezawa-shi, Yamagata 992-8510, Japan

[3] Advanced Institute for Materials Research, Tohoku University, 2-1-1 Katahira, Aoba-ku, Sendai-shi, Miyagi 980-8577, Japan

[4] Graduate School of Science and Engineering, Saitama University, 255 Shimo-Okubo, Sakura-ku, Saitama-shi, Saitama 338-8570, Japan

[5] Department of Electronics Engineering, Universitat Politècnica de Catalunya, Campus Nord, Building C4. C. Jordi Girona, 1-3, 08034 Barcelona, Catalunya, Spain

[6] Kansei Fukushi Research Institute, Tohoku Fukushi University, 6-149-1 Kunimi-ga-oka, Aoba-ku, Sendai-shi, Miyagi 989-3201, Japan

†These authors contributed equally to this work.
*Authors to whom correspondence should be addressed.
Email: ayumi.hirano.a5@tohoku.ac.jp

# Supplementary Results

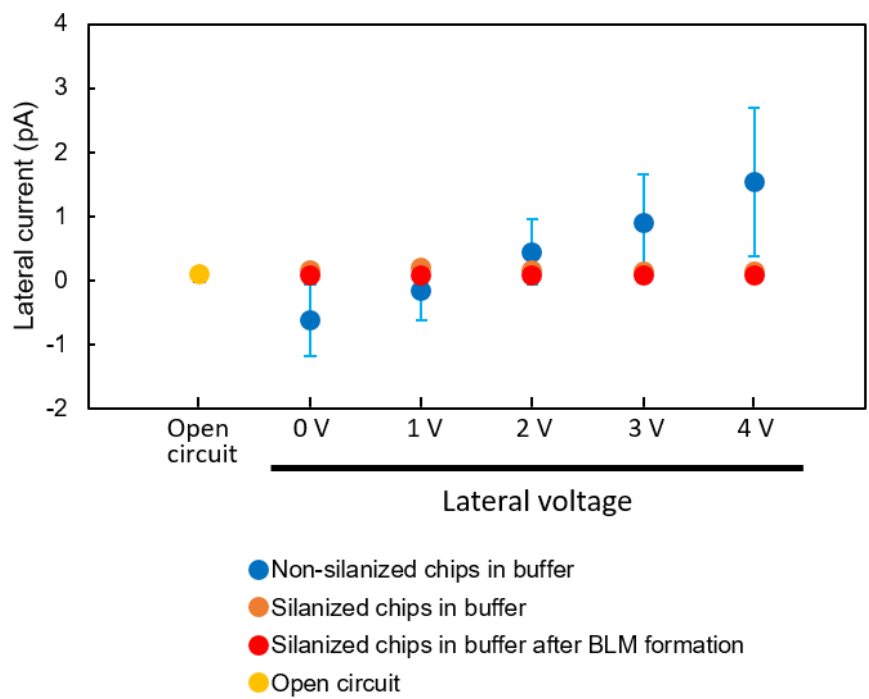

**Figure S1. Lateral currents under laterally applied voltages.**

The currents were measured with a KEITHLEY 2636B system sourcemeter which was connected in series with a DC power source and Teflon-based chips in buffer solutions. (blue) Bare Teflon-based chips. (orange) Teflon-based chips after being modified with PFDS. (red) PFDS-modified Teflon-based chips after BLM formation. (yellow) Current recorded when the sourcemeter was not connected with the Teflon-based chips nor the DC power source. The number of each trial was 5. The error bar is SEM.